\documentclass{cimento}

\usepackage{graphicx} 
\usepackage[]{units}
\usepackage{acronym}
\usepackage{upgreek}
\usepackage{caption}
\usepackage{subcaption}
\usepackage{ragged2e}
\usepackage{multirow}
\usepackage{float} 
\usepackage{booktabs}

\title{Measurement of the cosmic ray flux by an ArduSiPM-based muon telescope in the framework of the Lab2Go project.}
\author{V.~Agostini\from{ins:Pietrobono}, B.~Arcese\from{ins:Pietrobono}, N.~Ascani\from{ins:Pietrobono}, P.~Astone\from{ins:Roma1}, V.~Bocci\from{ins:Roma1}, S.~Caperna\from{ins:Pietrobono}, F.~Casaburo\from{ins:Sapienza}\from{ins:Roma1}\thanks{Corresponding author.}\thanks{Currently at Dipartimento di Fisica, Universit\`a di Genova and INFN- Sezione di Genova}, A.~Cerica\from{ins:Pietrobono}, C.~DAuria\from{ins:Pietrobono}, G.~De Bonis\from{ins:Roma1},  D.~Deda\from{ins:Pietrobono}, F.~Di Mauro\from{ins:Pietrobono}, A.~Di Vico\from{ins:Pietrobono}, R.~Faccini\from{ins:Sapienza}\from{ins:Roma1}, L.~Frasca\from{ins:Pietrobono}, G.~Galuppi\from{ins:Pietrobono}, G.~Giovannetti\from{ins:Pietrobono}, F.~Iacoangeli\from{ins:Roma1}, G.~Ludovici\from{ins:Pietrobono}, L.~Martone\from{ins:Pietrobono}, B.~Marucci\from{ins:Pietrobono}, L.~Mizzoni\from{ins:Pietrobono}, A.~Moriconi\from{ins:Pietrobono},
G.~Organtini\from{ins:Sapienza}\from{ins:Roma1}, F.~Piacentini\from{ins:Sapienza}\from{ins:Roma1}, 
A.~Pietrobono\from{ins:Pietrobono}, F.~Pongelli\from{ins:Pietrobono},
F.~Severa\from{ins:Pietrobono} \atque D.~Vona\from{ins:Pietrobono}}

\instlist{\inst{ins:Pietrobono} Liceo Scientifico "L. Pietrobono" - Alatri, Italy
\inst{ins:Sapienza} Dipartimento di Fisica, Universit\`a di Roma - Roma, Italy
\inst{ins:Roma1} INFN, Sezione di Roma - Roma, Italy
  }

\begin{document}

\maketitle

\begin{abstract}
Within \ac{INFN} outreach activities, the Lab2Go project is of great significance. Its goal is involving high school teachers and students in several laboratory activities, aiming at increasing the weight of experimental contents in teaching and learning. In this article we present the measurement, carried out in the framework of the Lab2Go project, of the cosmic muon flux made by an ArduSiPM-based muon telescope.

\end{abstract}

\section{Introduction}
\label{Sec:Introduction}

At the end of the XVIII century, the spontaneous discharge of electroscopes was observed \cite{Kir_ly_2013}. After R{\"o}ntgen's X-rays discovery \cite{Frankel1996}, it was proposed that the leakage of electric charge was due to the ionization caused by radioactivity coming from the underground. 
This hypothesis was dismissed by Pacini, who showed that the radiation under the sea was significantly lower than at the surface \cite{ArticoloPacini}. In 1912, during a balloon flight up to an altitude of $\unit[5200]{m}$, Hess observed an increasing of the radiation with the altitude \cite{https://doi.org/10.48550/arxiv.1808.02927}, proving that it comes from the space. Later, Millikan named this radiation \acp{CR} \cite{Bonolis_2011}. 

When \acp{CR} enter the atmosphere, they interact producing a particle showers called \ac{EAS}, that can be divided in three components: hadronic, electromagnetic and muonic \cite{Andreas_Haungs_2003}. The latter, due to the low interaction between muons and the atmosphere, can be detected at sea level. Here, muon flux is approximately of $\unitfrac[1]{\mu^{\pm}}{\unitfrac{min}{cm^{2}}}$ \cite{AUTRAN201877}, but it depends on the Zenith angle $\theta$ and, for $\theta<75{^\circ}$, it is:

\begin{equation}
\unitfrac[F\left(\theta\right)=F_{0}\cos^{2}\theta]{counts}{\nicefrac{s}{m^{2}}}
\label{eq:FluxLaw}
\end{equation}

being $F_0$ the flux at $ \theta=0$ \cite{Bektasoglu:2013eda}. Experimentally, we can measure it rotating the detector, and thus verifying the expected formal relation between measured flux and angle. In this article, we present the measurement of the cosmic muon flux made by an ArduSiPM-based muon telescope in the framework of the Lab2Go project \cite{ArticoloLab2Go} at Liceo "L. Pietrobono" in Alatri (Italy). Despite our experimental setup doesn't allow to estimate the electron contamination, due to the higher interaction of electrons in atmosphere, it is negligible for our educational purpose.

\section{Experimental setup and procedure}
\begin{figure}[htbp]
\begin{subfigure}{.5\textwidth}
  \centering
  \includegraphics[height=.31\linewidth]{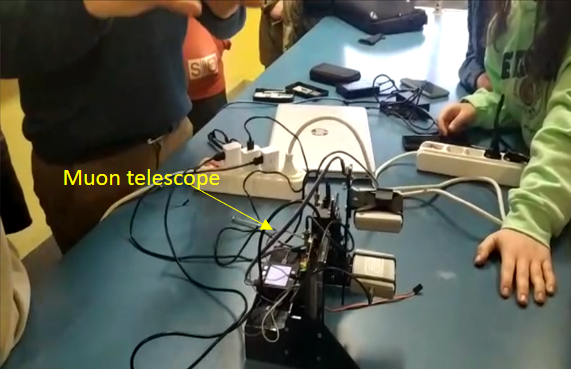}  
  \caption{}
  \label{fig:muontelescope}
\end{subfigure}
\begin{subfigure}{.5\textwidth}
  \centering
  \includegraphics[height=.31\linewidth]{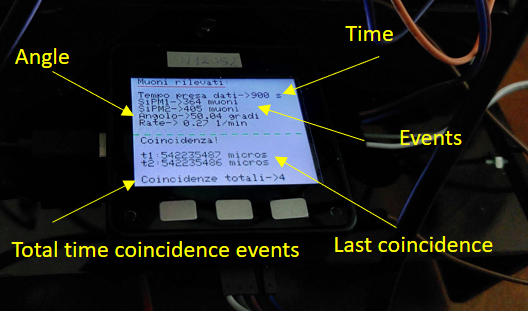}  
  \caption{}
  \label{fig:m5stack}
\end{subfigure}
\caption{(\subref{fig:muontelescope}) Muon telescope during the measurement at Liceo "L. Pietrobono". (\subref{fig:m5stack}) Data read by the M5Stack. }
\label{fig:Setup}
\end{figure}

 The experimental setup (Fig. \ref{fig:muontelescope}) consists of a muon telescope made of two ArduSiPMs. ArduSiPM is a transportable particle detector \cite{Ardusipm} created and developed by \ac{INFN} Roma, constituted by an electronic shield connected to an Arduino DUE \cite{ArduinoSite}, a $\unit[5]{cm}\times\unit[5]{cm}\times\unit[5]{mm}$ scintillator and a \ac{SiPM}. To build up the muon telescope, the two ArduSiPMs are placed on a mechanical structure that hosts the scinitllators as well, and is free to rotate. The distance between scintillators is $\unit[7.8]{cm}$, corresponding to an acceptance of 0.16 \cite{TesiCarlotta}. To automatize the angle measurement, the inclination is derived by $\vec{g}$ components measured by an accelerometer at reast connected to the rotating mechanical structure.
 

 The number of events measured by each one of the ArduSiPMs and by both of them (time width= $\unit[10]{\mu s}$), the acquisition time and the angle are read by an M5Stack (Fig. \ref{fig:m5stack}) \cite{M5Stack}. 

\section{Data analysis and results}
The number of coincidences per unit time was measured at different angles as $\frac{N\left(\theta\right)}{T}$, being $N\left(\theta\right)$ the number of events detected by both the ArduSiPM and $T$ the acquisition time. Due to the low rate of detected muons (it depends on the acceptance), long acquisition time was needed for each angle, from approximately 1h at $\theta=0$ up to approximately 2h at $\theta=60^{\circ}$. The uncertainty on $N\left(\theta\right)$ has been evaluated assuming a Gaussian uncertainty when counts were $>$ 30, and a Poisson uncertainty otherwise.

\begin{figure}[H]
\centering
\includegraphics[width=1.8in]{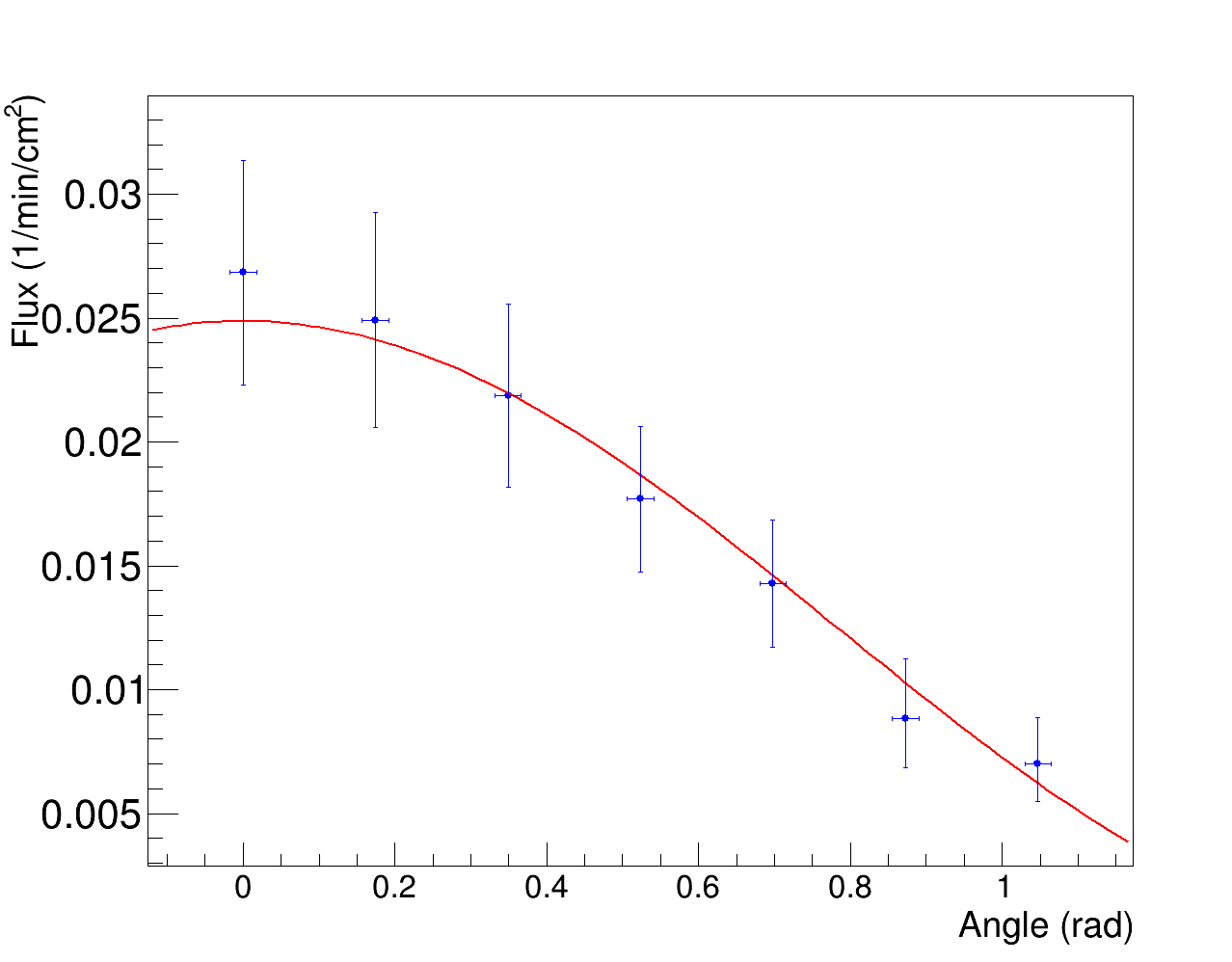}\DeclareGraphicsExtensions.
\caption{Flux interpolation. }
\label{fig:fitflux}
\end{figure}

 Being $A$ the area of the scintillator, values of $F\left(\theta\right)=\frac{N\left(\theta\right)}{TA}$ for several angles have been interpolated (Fig. \ref{fig:fitflux}) with the function Eq. \ref{eq:FluxLaw}, resulting in $F_{0_{esti}}=\unitfrac[\left(2.49\pm0.18\right)\cdot10^{-2}]{counts}{\unitfrac{min}{cm^{2}}}$ as derived from the fit. Despite the large uncertainties due to poor statistics (few tens of detected muons), the relation in Eq. \ref{eq:FluxLaw} is qualitatively verified as shown in Fig. \ref{fig:fitflux}.

In addition, to further stress out the need for a quantitative validation and to allow the students to acquire the concept of agreement between measurements, we evaluated a test function. In particular, we started assuming no-correlation between the estimated $F_{0_{esti}}$ and measured $F_{0_{meas}}=\unitfrac[\left(2.68\pm0.45\right)\cdot10^{-2}]{counts}{\unitfrac{min}{cm^{2}}}$ flux parameters (zero-hypothesis) and we consider $t=\frac{\left|F_{0_{esti}}-F_{0_{meas}}\right|}{\sqrt{\sigma_{F_{0_{esti}}}^{2}+\sigma_{F_{0_{meas}}}^{2}}}$, whose result ($t=0.4$) has been used to estimate a one-sided gaussian p-value. We got p-value=0.7 ($>0.05$, then verified at 95\% C.L.) and an agreement within $1\sigma$.

\section{Students' satisfaction}
\begin{figure}[H]
\centering
\includegraphics[width=1.8in]{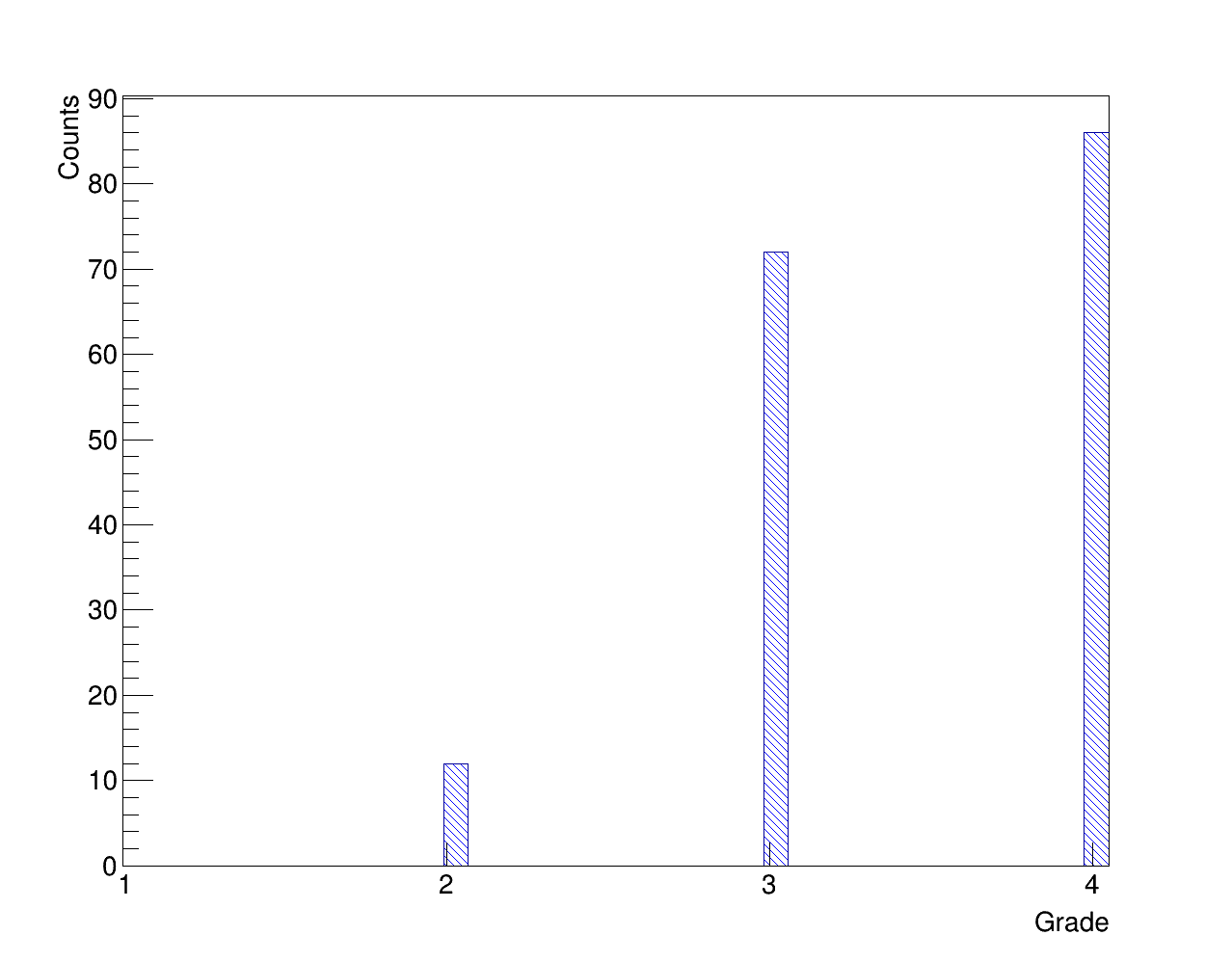}\DeclareGraphicsExtensions.
\caption{Results of satisfaction questionnaires.}
\label{fig:questionnaire}
\end{figure}

At the end of the Lab2Go course, an anonymous satisfaction questionnaire has been proposed to each one of 17 students (10 male and 7 female) attending Lab2go, in order to evaluate their level of satisfaction concerning the lecture topics, their engagement, the quality of teaching and the overall project. For each of the 10 questions, students (attending the last three years of Liceo Scientifico) could express their impression, from a minimum of 1 (unsatisfactory) to a maximum of 4 (very  satisfactory). The results of the total 170 answers (17 students * 10 questions) have been summarised in an histogram (Fig. \ref{fig:questionnaire}) resulting in a mean grade $\mu=3.435\pm0.048$.

\section{Conclusions}
In the framework of Lab2Go, the measurement of the cosmic muon flux by an ArduSiPM-based muon telescope has been proposed to students of Liceo "L. Pietrobono" in Alatri. 
Thanks to this experiment we had the possibility to introduce many topics, as \acp{CR}, particle physics, muons, antimatter, special relativity, particle detectors and time coincidence, going far beyond the topics commonly taught in physics lectures at high schools. 
Moreover, students learned how to process, interpolate and show data, taking into account the uncertainties and estimating the agreement between measurements. A 
 satisfaction questionnaire proposed to students at the end of the course reported a positive evaluation of the proposed activity and, in general, of Lab2Go overall.

\acknowledgments
The authors acknowledge Mauro Mancini, Francesco Safai Therani, and \ac{CC3M}-\ac{INFN}.

\bibliographystyle{ieeetr}
\bibliography{sample}

\acrodef{CR}[CR]{Cosmic Ray}
\acrodef{CC3M}[CC3M]{Comitato di Coordinamento III missione}
\acrodef{EAS}[EAS]{Extensive Air Showers}
\acrodef{INFN}[INFN]{Istituto Nazionale di Fisica Nucleare}
\acrodef{SiPM}[SiPM]{Silicon PhotoMultiplier}


\end{document}